\documentclass[]{emulateapj}

\usepackage{graphicx}				

\usepackage[hang,flushmargin]{footmisc}

\shortauthors{Bedell et al.}
\shorttitle{Highest-Precision Abundances}

\begin{document}
\DeclareGraphicsExtensions{.eps,.png}

\title{Stellar Chemical Abundances: In Pursuit of the Highest Achievable Precision}

\author{Megan Bedell\altaffilmark{1},
Jorge Mel\'{e}ndez\altaffilmark{2},
Jacob L. Bean\altaffilmark{1},
Ivan Ram\'{i}rez\altaffilmark{3},
Paulo Leite\altaffilmark{2},
Martin Asplund\altaffilmark{4}}

\email{E-mail: mbedell@oddjob.uchicago.edu}

\altaffiltext{1}{Department of Astronomy and Astrophysics, University of Chicago, 5640 S. Ellis Ave, Chicago, IL 60637, USA}
\altaffiltext{2}{Departamento de Astronomia do IAG/USP, Universidade de
S\~{a}o Paulo, Rua do Mat\~{a}o 1226, Cidade Universit\'{a}ria, 05508-900 S\~{a}o Paulo, SP, Brazil}
\altaffiltext{3}{McDonald Observatory and Department of Astronomy, University of Texas at Austin, USA}
\altaffiltext{4}{Research School of Astronomy and Astrophysics, The Australian National University, Cotter Road, Weston, ACT 2611, Australia}


\begin{abstract}

The achievable level of precision on photospheric abundances of stars is a major limiting factor on investigations of exoplanet host star characteristics, the chemical histories of star clusters, and the evolution of the Milky Way and other galaxies.  While model-induced errors can be minimized through the differential analysis of spectrally similar stars, the maximum achievable precision of this technique has been debated.  As a test, we derive differential abundances of 19 elements from high-quality asteroid-reflected solar spectra taken using a variety of instruments and conditions.  We treat the solar spectra as being from unknown stars and use the resulting differential abundances, which are expected to be zero, as a diagnostic of the error in our measurements.  Our results indicate that the relative resolution of the target and reference spectra is a major consideration, with use of different instruments to obtain the two spectra leading to errors up to 0.04 dex.  Use of the same instrument at different epochs for the two spectra has a much smaller effect ($\sim$0.007 dex).  The asteroid used to obtain the solar standard also has a negligible effect ($\sim$0.006 dex).  Assuming that systematic errors from the stellar model atmospheres have been minimized, as in the case of solar twins, we confirm that differential chemical abundances can be obtained at sub-0.01 dex precision with due care in the observations, data reduction and abundance analysis.

\end{abstract}

\section{Introduction}

The derivation of photospheric chemical abundances from high-resolution stellar spectra is a key tool in studies of galactic chemical enrichment, planet formation, and the physics of stars.  Recent advances in these fields have relied on trends in chemical abundance data at the level of 0.01 dex, a previously unattainable precision \citep[e.g.][]{Melendez2009,Melendez2014,Melendez2014b,Ramirez2009,Ramirez2014}.  

Differential comparison of stars at this level of abundance precision has major implications for investigations of planet hosting stars' compositions, a long-standing topic of research in the field of exoplanets \citep[e.g.][]{Gonzalez1997, Santos2004, Fischer2005}.   Specific investigations have dealt with the prospect of Li as a marker of planet presence \citep[e.g.][]{King1997, Israelian2009, Delgado-Mena2014} or stellar age \citep[e.g.][]{Baumann2010,Monroe2013,Melendez2014}; refractory element abundances in planet host stars as an indicator of planet formation processes \citep[e.g.][]{Brugamyer2011,Carter-Bond2012}; comparisons of the compositions of host stars and their planets \citep[e.g.][]{Petigura2011,Teske2014}; and trends in abundance with condensation temperature as a potential signature of planet formation \citep[e.g.][]{Melendez2009, Schuler2011b, Ramirez2013, GonzalezHernandez2013, TucciMaia2014}.  With a precision of 0.01 dex, one can potentially detect differences in composition corresponding to just a few Earth masses of refractory materials \citep{Chambers2010}.  We also note the recent work on abundance analyses of white dwarfs revealing the composition of planets accreted onto the star \citep[e.g.][]{Farihi2013}.

High-precision stellar abundances are also crucial to our understanding of the evolution of star clusters and galaxies.  Comparisons of stars within a single cluster can shed new light on its history of nucleosynthetic processes \citep[e.g.][]{Yong2013}.  Comparisons of more distant stars can help to reconstruct the dynamical history of the local Milky Way through chemical tagging \citep[e.g.][]{Freeman2002, Bergemann2014, Ramirez2014b}.  With better precisions on stellar abundances, these histories can be inferred at an unprecedented level of detail \citep{Lindegren2013}.

Past analyses have often adopted a lower-limit error of 0.02-0.05 dex for abundance measurements,\footnote{Throughout this paper we quote abundance of the element X relative to hydrogen in the standard form of [X/H] = $A_{X, \star} - A_{X, \odot}$, with $A_{X} = log(n_X/n_H) + 12$ where $n_X$ is the number density of element X.} considering precisions at or below 0.01 dex to be impossible with current capabilities \citep{Asplund2009}.  This assumed uncertainty comes in part from awareness of potential systematic errors stemming from the model atmospheres employed in the analysis.  Approximations such as one-dimensionality and assumption of local thermal equilibrium, as well as uncertain treatments of turbulent behaviors in the stellar photosphere, contribute biases to the models.  These biases can manifest as false trends in abundance with stellar parameters at or above the level of 0.05 dex \citep[e.g.][]{Valenti2005, Asplund2005}.  For high-resolution, high-signal-to-noise data, this model-induced error is usually the dominant source of uncertainty on derived abundances.  

Model-induced errors can be minimized through careful choice of the sample of stars to be analyzed.  By comparing stars within a narrow range of stellar parameters and performing all analyses in a strictly differential sense, the impact of all unknown systematic errors can be characterized by the reference and subtracted out.  This approach is especially valuable for solar twin stars, where the stellar parameters of the reference star (the Sun) are independently known \citep[e.g.][]{Melendez2009, Takeda2009, Melendez2012, Monroe2013}.

With model-dependent errors minimized, it is critical to give thorough consideration to other sources of error in order to determine the achievable precision of the technique.  One potential error source is the Sun itself, when assumed to be a constant reference which yields the same intrinsic spectrum in all observations.  This assumption has been investigated by \citet{Kiselman2011}, who check for latitudinal variations over the surface of the Sun and find that any latitude-dependent effects should manifest below the level of 0.005 dex.  In this paper, we additionally search for potential spectral distortions due to the use of asteroids as reflectors.

Other fundamental limitations on the maximum achievable precision come from the quality of the spectra used.  Since the abundance analysis technique relies on differential measurements between the target and reference spectra, a variation between the two spectra arising from their being taken at different times or with different instruments could introduce errors.  We investigate this possibility using multiple solar spectra taken under different conditions.

In this paper we present a detailed error budget for high-precision stellar chemical abundance analyses.  Using high-resolution, high-signal-to-noise spectra of reflected sunlight from asteroids, we examine several contributors to the total abundance uncertainty. We analyze factors including time-dependent instrumental and atmospheric variability, use of different reference asteroids, use of different instruments, errors in the line equivalent width measurements, and the uncertainty in stellar parameters.

\bigskip
\section{Observations}

\begin{table}
\caption{Summary of observations.}
\label{obs}
\centering 
\begin{tabular}{llccc} 
\hline    
\hline 
{Name}& Date & $m_V$ & SNR \footnotemark[1] & AM \footnotemark[2]  \\
\hline
Vesta (ESPaDOnS) & 2013 Mar 04  & 8.0  & 691 & 1.05 \\
Ceres (ESPaDOnS) & 2013 Mar 04 & 8.3  & 663 & 1.01 \\
Vesta (MIKE, 1) &  2011 June 24 & 6.4 &  730 & 1.05 \\
Vesta (MIKE, 2) &  2011 Sept 09 & 6.4 &  764 & 1.00 \\
Iris (MIKE) & 2011 Jan 04 & 8.2 &  588 & 1.33 \\
\hline       
\end{tabular}
\footnotetext[1]{Signal-to-noise ratio at $\sim$ 6000 \r{A}.}
\footnotetext[2]{Airmass at the start of observation.}
\end{table}

The five solar spectra used in this analysis were obtained with very high resolution and signal-to-noise ratios (SNRs) characteristic of data used in past stellar abundance analyses.  Two spectra were taken with the Echelle SpectroPolarimetric Device for the Observation of Stars (ESPaDOnS) instrument \citep{Donati2003} at the 3.6 meter Canada-France-Hawaii telescope on the night of 2013 March 4.  The asteroids Ceres and Vesta were each observed in ``star only" mode at a spectral resolving power R = 81000.  The spectra have complete coverage over a wavelength range of 380 to 880 nm.  Observing conditions for the two spectra were made as identical as possible by observing Vesta immediately after Ceres and at a similar airmass.  Observation details are listed in Table \ref{obs}.  The spectra were reduced with the Upena pipeline,\footnote{\url{http://www.cfht.hawaii.edu/Instruments/Upena/index.html}} which employs the Libre-ESpRIT package to reduce and optimally extract each order, perform wavelength calibration, and apply an approximate continuum normalization \citep{Donati1997}. Further normalization was performed using a polynomial fit to the spectrum in 100-\r{A} chunks, with polynomial orders ranging from 2 to 7.

The remaining three solar spectra were taken with the Magellan Inamori Kyocera Echelle (MIKE) spectrograph \citep{Bernstein2003} at the 6.5 meter Magellan Clay telescope.  The asteroid Vesta was observed twice and Iris was observed once during three separate observing runs spanning January to September of 2011.  All observations were carried out in MIKE's standard setup with the 0.35 arcsecond width slit, giving a spectral resolving power of R = 83000 on the blue CCD and 65000 on the red CCD.  Further details of the observations are in Table \ref{obs}.  The MIKE spectra were processed with the CarnegiePython MIKE pipeline\footnote{\url{http://code.obs.carnegiescience.edu/mike}} and barycentric corrections were applied with IRAF's\footnote{IRAF is distributed by the National Optical Astronomy Observatory, which is operated by the Association of Universities for Research in Astronomy (AURA) under cooperative agreement with the National Science Foundation.}  \textit{dopcor} and \textit{rvcor} tasks.  Each spectral order was trimmed of $\sim$ 100 pixels at each end and continuum normalizations were performed using 12th order polynomial fits to the upper envelopes of the data.  Furthermore, the 5 reddest orders and the 19 bluest orders were discarded due to unreliable continuum normalization.  The orders were merged into a single one-dimensional spectrum using IRAF's \textit{scombine} task.  The resulting reduced spectra have complete wavelength coverage between 400 and 800 nm.

\bigskip
\section{Abundance Analysis}

\subsection{Line Measurements}

\begin{table*}
\caption{Line List.}
\label{linelist}
\centering 
\begin{tabular}{ccccccccc} 
\hline
\hline
{Wavelength} & Species & EP & log($gf$) & Vesta (ESPaDOnS) & Ceres (ESPaDOnS) & Vesta (MIKE, 1) & Vesta (MIKE, 2) & Iris (MIKE) \\
 (\r{A}) &  & (eV) &  & EW (m\r{A}) & EW (m\r{A}) & EW (m\r{A}) & EW (m\r{A}) & EW (m\r{A}) \\
\hline
5052.17 &  6.0 & 7.68 & -1.24 &  33.6 &  34.2 & & & \\
5380.34 &  6.0 & 7.68 & -1.57 &  18.3 &  19.0 &  21.6 &  21.1 &  21.6 \\
6587.61 &  6.0 & 8.54 & -1.05 &  13.4 &  13.4 &  14.7 &  13.9 &  14.2 \\
7111.47 &  6.0 & 8.64 & -1.07 &   9.4 &   9.5 & & & \\
7113.18 &  6.0 & 8.65 & -0.76 &  19.7 &  19.1 &  23.1 &  22.7 &  22.9 \\
  &  & & & & $\vdots$   & &  &  \\
\hline       
\end{tabular}
\tablecomments{Table \ref{linelist} is published in its entirety in the electronic edition of ApJ. A portion is shown here for guidance regarding its form and content.}

\end{table*}

For the analysis of chemical abundances, we employed a line list consisting of 97 Fe I lines, 18 Fe II lines, and 167 lines of other elements (C, O, Na, Mg, Al, Si, S, K, Ca, Sc, Ti, V, Cr, Mn, Co, Ni, Cu, and Zn).  The line list was based on the list employed in \citet{Melendez2014b}, with minor modifications for differing wavelength coverage and telluric line presence in the ESPaDOnS and MIKE spectra.  Lines were selected for these analyses with a preference for unsaturated lines with minimal blending.  The full line list is presented in Table \ref{linelist}.  Atomic parameters were taken from laboratory transition probabilities when available and supplemented with theoretical or solar gf-values.  For the differential analysis technique employed the exact gf-values adopted are irrelevant, since they cancel out during calculation of the differential abundances.

We measured all line equivalent widths (EWs) by hand using the \textit{splot} task in IRAF to fit a Gaussian to each line.  If necessary, multiple Gaussians were fit in the case of a blend. Since the wavelength coverage of the MIKE and ESPaDOnS instruments are different, spectra from each instrument were analyzed separately with slightly modified line lists.  Careful attention was given to ensure that the same continuum region and wavelength interval were used to fit lines across all spectra from the same instrument.

\begin{figure}[h!]
\includegraphics{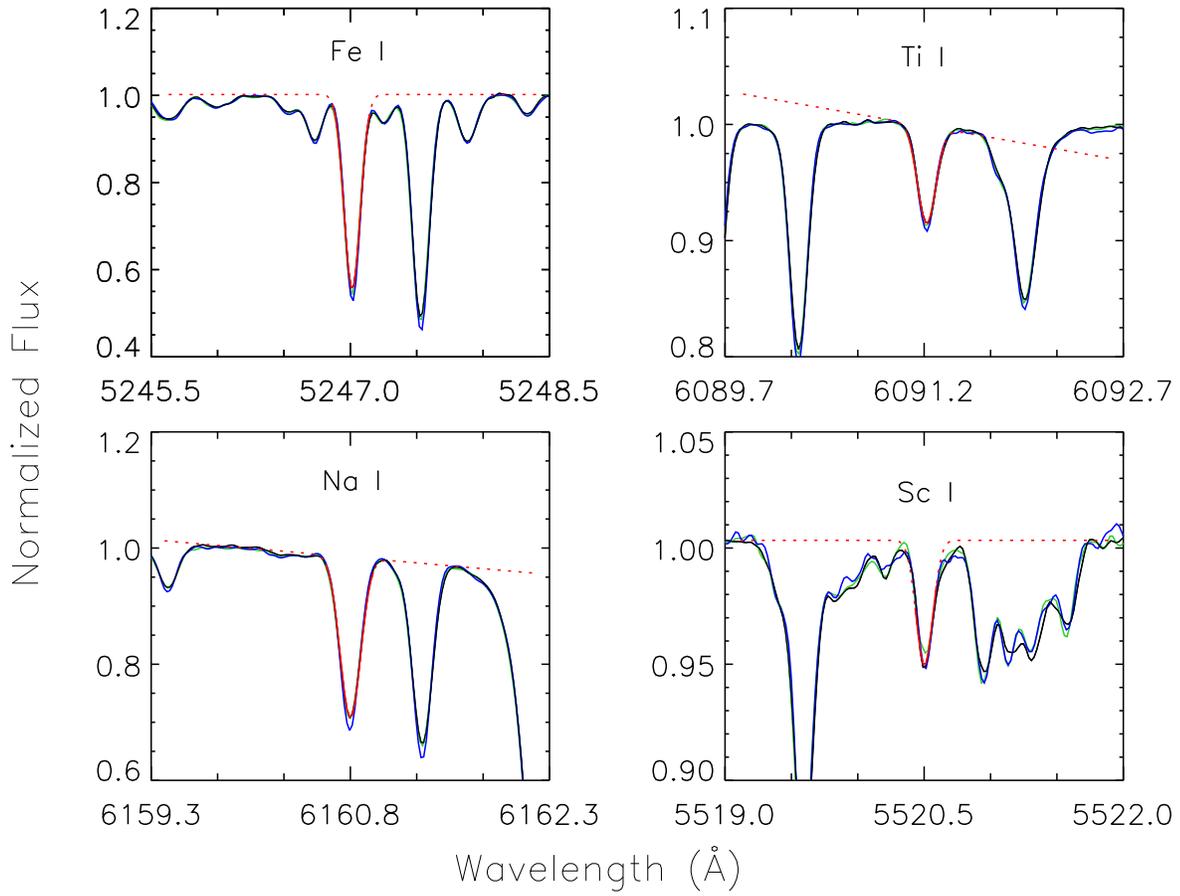}
\caption{Example spectral lines with MIKE Vesta 1 data plotted in black and Gaussian fits to the data plotted in red, with the estimated local continuum extrapolated as a dotted line.  MIKE Vesta 2 and Iris spectra are plotted in green and blue respectively for comparison.  In some cases, the local pseudo-continuum selected for equivalent width measurements is noticeably different from the true stellar continuum.  Choice of pseudo-continuum is made to minimize the potential effects of nearby unresolved features or line wings which may be slightly blended with the line being measured, introducing a local slope or offset.}
\label{fig:ews}
\end{figure}

A strictly accurate absolute measurement of equivalent width would depend on finding the true spectral continuum.  We chose instead to use a ``pseudo-continuum" approach which employs whichever point(s) in the immediate vicinity of the line appear most constant across the multiple spectra being measured. The aim of this approach is to measure EWs with the highest possible precision or consistency across multiple spectra, minimizing the impact of nearby features on the line in question.  Since our line list consists of lines in the linear region of the curve of growth, a small discrepancy between the measured EWs and the true EW has an insignificant impact on the resulting differential abundance as long as this discrepancy is equally present in the target and reference spectrum measurements.  The pseudo-continuum approach is especially valuable in the case of a crowded spectral region in which the spectrum has a local slope across the measured line due to the wings of neighboring lines or unresolved broad features (see the examples of Ti I and Na I in Figure \ref{fig:ews}).  The use of very local pseudo-continuum points is not always the best option, however.  In some cases, such a choice would result in a pseudo-continuum level too low to accurately fit a Gaussian profile to the line (as in the example of Fe I in Figure \ref{fig:ews}).  In other cases, any local slope is minimal enough compared to the level of noise in the spectrum that choosing two local points rather than a broader swath of nearby continuum would only add noise to the measurement (as in the example of Sc I in Figure \ref{fig:ews}).  The optimal continuum or pseudo-continuum choice is largely a judgment call made on a line-by-line basis.

We carried out tests to evaluate the validity of this measurement technique compared to more classical methods, measuring a set of 40 Fe I lines in the MIKE Vesta 1 and Vesta 2 spectra using several strategies and comparing the scatter in the resulting differential abundances.  The measurement technique used in this work, with a combination of pseudo-continua and ``true" continua chosen on a line-by-line basis, yields a resulting scatter in abundances measured by the standard error on the mean of 0.0023 dex.  This value was revised down to 0.0016 dex when the five largest outliers were remeasured with slightly different continuum choice.  In contrast, adopting a strict ``true" continuum for all lines yields a scatter of 0.0024 dex, and adopting the two neighboring points to the line as a peudo-continuum in every case yields a scatter of 0.0064 dex.  For further comparison we used the automated equivalent width measuring code ARES \citep{Sousa2007}, which gave a scatter of 0.0035 dex.

The subjectivity of an individual's choice of pseudo-continuum level could potentially bias results, especially in the case of elements for which only a few lines were measured.  To mitigate this effect, seven elements (C, O, Mg, Al, S, Sc, and Cu) were independently analyzed by two of us in a blind test.  The resulting abundances for the re-measured elements were generally within the one-sigma statistical error bars of the original abundances.

\subsection{Stellar Parameter Determination}

A critical first step in abundance determination is the derivation of correct stellar parameters (effective temperature $T_{\textrm{eff}}$, surface gravity $log(g)$, microturbulent velocity $v_t$, and metallicity [M/H]) for the target star.  Although all spectra used in this work are solar, we treat the target spectra as unknown stars and determine their parameters using spectral iron lines as we would for any other star.  Measured Fe I and Fe II line EWs were converted to abundances using the 2002 version of MOOG \footnote{\url{http://www.as.utexas.edu/~chris/moog.html}} with the \textit{abfind} driver, which employs a curve-of-growth method.    We used Kurucz ODFNEW model atmospheres\footnote{\url{http://kurucz.harvard.edu/grids.html}} and linearly interpolated between grid points to achieve the required resolution in parameter space.  Stellar parameters for each target spectrum were found through a differential approach with respect to a reference spectrum, arbitrarily taken as the Vesta 1 spectrum for the MIKE data and Vesta for the ESPaDOnS data.  We fix the parameters of the reference spectrum to the nominal solar values, $T_{\textrm{eff}}$ = 5777 K, $log(g)$ = 4.44 dex, and [M/H] = 0.0, and find the optimal microturbulence value by minimizing the trend between derived Fe I abundance and reduced equivalent width.  Throughout all analyses, [M/H] was assumed to be equal to [Fe/H].  

We then determine the stellar parameters for other spectra by computing iron abundances for the target star and the reference and imposing requirements on the \textit{differential} abundances (target - reference), as described in \citet{Melendez2014}.  These requirements consist of minimal slopes in Fe I abundance with excitation potential (primarily sensitive to the assumed model $T_{\textrm{eff}}$) and with reduced equivalent width (primarily sensitive to the assumed microturbulence value), minimal difference between the derived abundances of Fe I and Fe II (primarily sensitive to the assumed surface gravity), and equivalence between the input metallicity on the stellar atmosphere model and the output Fe abundance.  Our key assumption in using the differential abundances is that any systematic model errors manifesting in the reference star's abundances (e.g. a slope in Fe I abundance with excitation potential when using a model with the nominal solar effective temperature) are identical for the target star.  These systematic errors should therefore be subtracted out of the abundances before equilibrium conditions are considered.  Since all spectra considered in this paper are solar, this assumption is trivially valid.  In a more general sense the assumption should hold for target stars which are sufficiently close to the chosen reference star in parameter space; see for example the evaluation of potential systematics in stellar parameters for a solar twin sample in \citet{Ramirez2014}.

Uncertainties on our derived stellar parameters were estimated using the method described in \citet{Epstein2010} and \citet{Bensby2014}.  In brief, the observational errors on the quantities which were minimized to find the optimal parameter solution are propagated while accounting for the dependence of each stellar parameter on the others.  For the metallicity parameter, the uncertainty was taken as this formal parameter uncertainty added in quadrature with the line-to-line scatter on [Fe/H] derived in the spectral analysis.  The parameter solutions and uncertainties for each spectrum analyzed are shown in Table \ref{parameters}.

\subsection{Abundance Measurements}

After the optimal model atmosphere was chosen, we determined abundances for all elements in the line list using MOOG.  For elements which were observed in multiple ionization states (Fe, Sc, Ti, and Cr), the final abundance was taken as the error-weighted average of the abundances from each ionization state.  Hyperfine structure corrections were applied for four elements (Cu, Co, Mn, and V) using MOOG's \textit{blends} driver.  All abundances used in this paper assume local thermal equilibrium (LTE).  Non-LTE corrections can be crucial for accurate absolute abundance values, but our approach aims only for high precision on the differential measurements, so that applying a similar non-LTE correction to both reference and target stellar abundances creates a very small change in the derived values.  Past solar twins work in \citet{Melendez2012} has shown that non-LTE corrections have a negligible effect (on order of 0.001 dex) on differential abundances for stars with extremely similar parameters to the reference solar spectrum.

\medskip
\section{Estimated Error Budget}

We consider several combinations of spectra which introduce various potential sources of error on the derived abundances, including time-dependent effects, use of different asteroids, and use of different instruments.  The scatter in abundance derived from lines of the same element gives an estimate of the level of random error associated with these uncertainty sources.  We also use the deviation of the derived abundances from the expected solar values ([X/H] $\equiv$ 0.0 dex) as an estimate of more systematic uncertainty which may be underestimated in the scatter-based error bars.  These statistics are compared to the expected errors from formal uncertainties on the line equivalent width measurements and on the derived stellar parameters.

\bigskip
\subsection{Observed Errors}

\subsubsection{Time-Dependent Systematics}

Instrumental systematics like mechanical flexure and internal scattering of light may change considerably over time, so it is reasonable to expect that spectra taken with the same instrument at significantly different times would have some additional error due to varying spectral quality.  Additionally, the Earth's atmosphere is a major contributor to time-dependent spectral variations as weather evolves and as the target spectrum shifts with respect to telluric features due to the Doppler effect of the Earth's rotation.  Although we discard obviously telluric-contaminated lines from the list, small unresolved tellurics are still a concern.

\begin{figure}
\includegraphics[width=\columnwidth]{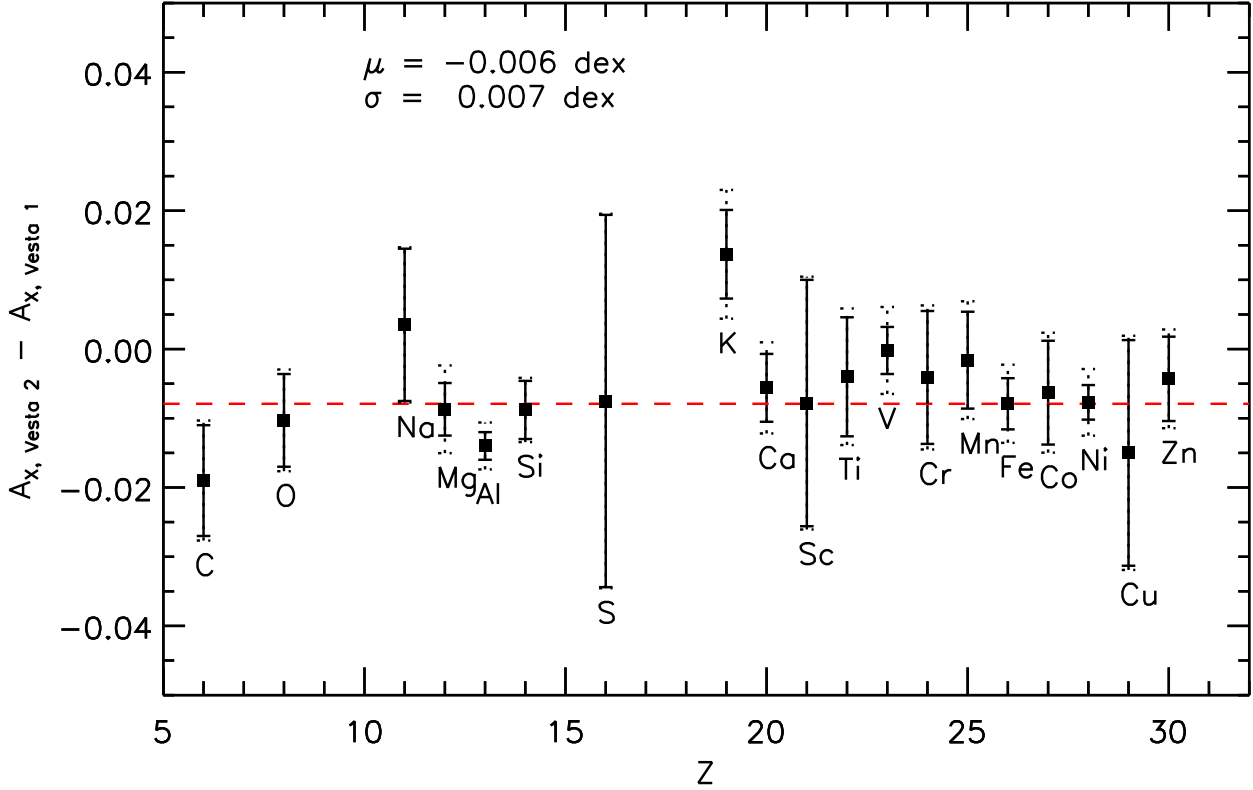}
\caption{Differential abundances for Vesta spectra taken at two epochs with the MIKE instrument.  Solid error bars represent the standard error on the mean differential abundance derived from the sample of lines measured.  Dotted error bars represent the ``total" error from adding statistical error and error from uncertainty on the stellar parameters in quadrature.   Red dashed line is at the level of the derived [Fe/H] abundance.}
\label{fig:vestamike}
\end{figure}

We quantify the error due to time-dependent systematics using two spectra of Vesta taken with the MIKE instrument on dates June 24, 2011 (``MIKE Vesta 1") and September 09, 2011 (``MIKE Vesta 2").   The derived stellar parameters for the Vesta 2 spectrum relative to the Vesta 1 spectrum are extremely close to the expected solar values, with $T_{\textrm{eff}}$ differing by 2 $\pm$ 5 K, $log(g)$ by 0.00 $\pm$ 0.01 dex, $v_t$ by 0.01 $\pm$ 0.01 km s$^{-1}$, and metallicity by -0.01 $\pm$ 0.01 dex (Table \ref{parameters}).  The derived [X/H] abundances are shifted down to a mean of -0.006 dex as a result of this metallicity value, but the standard deviation among [X/H] abundances is only 0.007 dex, indicating that the abundance of one measured element relative to another (e.g. [X/Fe]) can be considered reliable to below 0.01 dex precision (Figure \ref{fig:vestamike}, Table \ref{abund_vestamike}).

The most significant outlier from the mean is K ([K/H] - $\langle$[X/H]$\rangle$ = 0.020 dex, or 3.3$\sigma$ when using the line-to-line scatter as an error bar).  This is a less significant deviation when the error due to parameter uncertainties is taken into account (bringing the deviation down to 2.2$\sigma$; see section \ref{paramerrors}), but it is still an indication of the limitations of employing a short line list.  In our list, K is the only element for which only a single line was measured.  An estimate of ``line-to-line scatter" was obtained from remeasuring the line multiple times with slightly different but still acceptable continuum choices and taking the standard deviation of the resulting abundances, but this error estimate neglects any potential effects on the line such as an unresolved blend that will remain regardless of the local continuum choice.

In general, the consistency of abundances for all elements and their agreement with the expected solar values ([X/Fe] $\equiv$ 0.0) demonstrate that time-dependent systematics for the MIKE instrument over a timescale of months cause abundance errors well below the 0.01 dex level.

\subsubsection{Choice of Asteroid}

Differential abundance analyses frequently employ reference spectra of solar light reflected from the brightest asteroid at the time of observation, which can vary between different observing runs.  Thus a potential error in comparing stellar abundances derived from stars observed at different times could be varying properties of the reference solar spectrum depending on which asteroid was used for each star's reference.  It is generally expected that asteroid reflectance properties should have a negligible effect on the observed solar spectrum, since reflectance does not change significantly within wavelength ranges below a few hundred \r{A} \citep{Xu1995,Binzel1996, DeMeo2009}.  Nevertheless, the possibility of spectral variations from chemical activity such as water evaporation on Ceres \citep{Kuppers2014} makes the use of different asteroids a source of error worth investigating.

\begin{figure}
\includegraphics[width=\columnwidth]{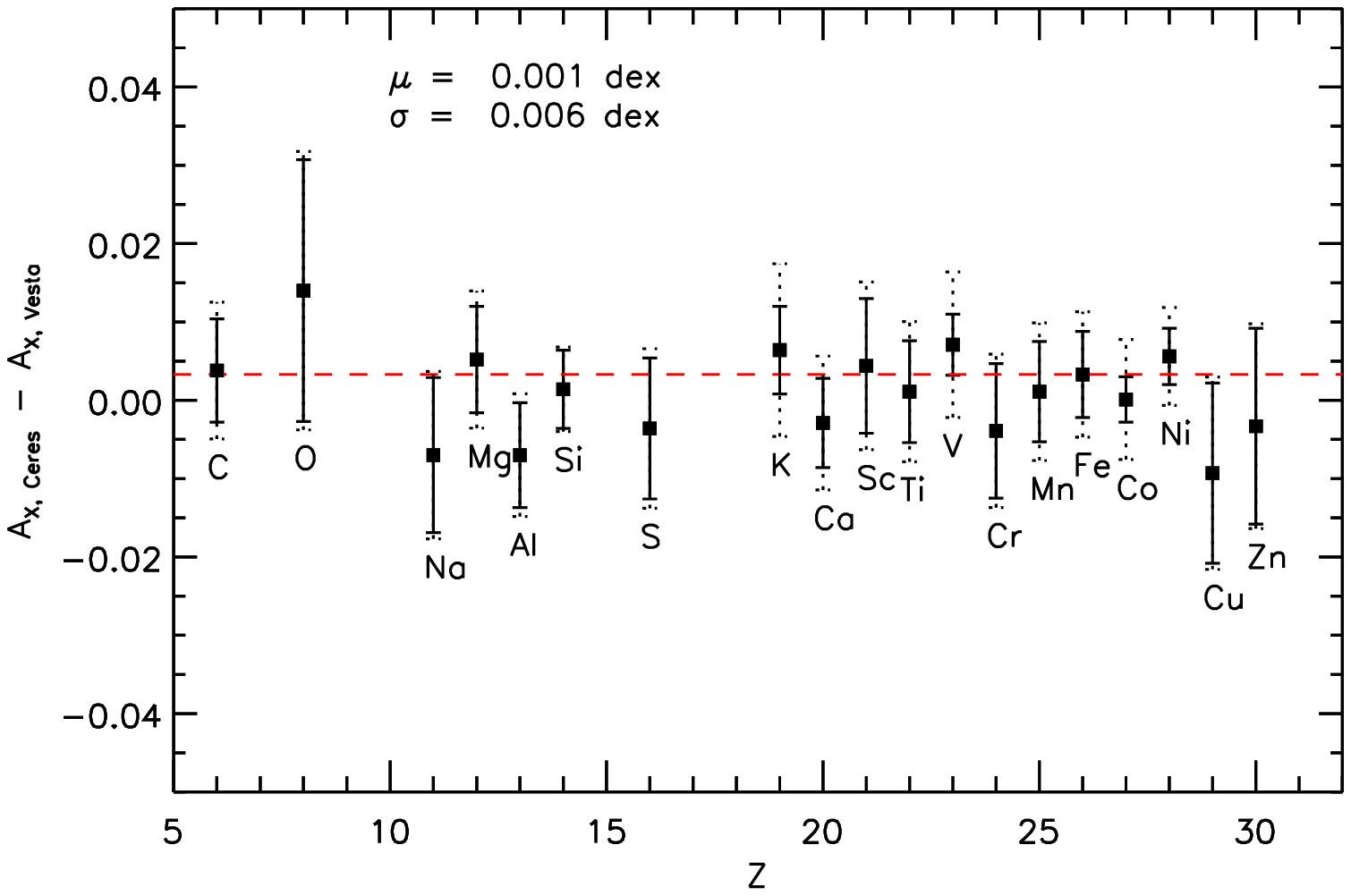}
\caption{Differential abundances for Ceres and Vesta data from the ESPaDOnS instrument.  Error bars and red line as in Figure \ref{fig:vestamike}.}
\label{fig:espadons}
\end{figure}

We investigate the possibility of asteroid-dependent errors using the ESPaDOnS spectra of Ceres and Vesta, which were observed very close in time and should have minimal time-dependent errors.  The asteroids Ceres and Vesta have significantly different reflectance properties \citep{DeMeo2009}, making them good test subjects.  The derived stellar parameters for Ceres with respect to Vesta are extremely accurate and of comparable precision to the MIKE pairs (Table \ref{parameters}).  The resulting [X/H] values have a standard deviation of 0.006 dex and a mean value of 0.001 dex with no major outliers (Figure \ref{fig:espadons}, Table \ref{abund_esp}).

\begin{figure}
\includegraphics[width=\columnwidth]{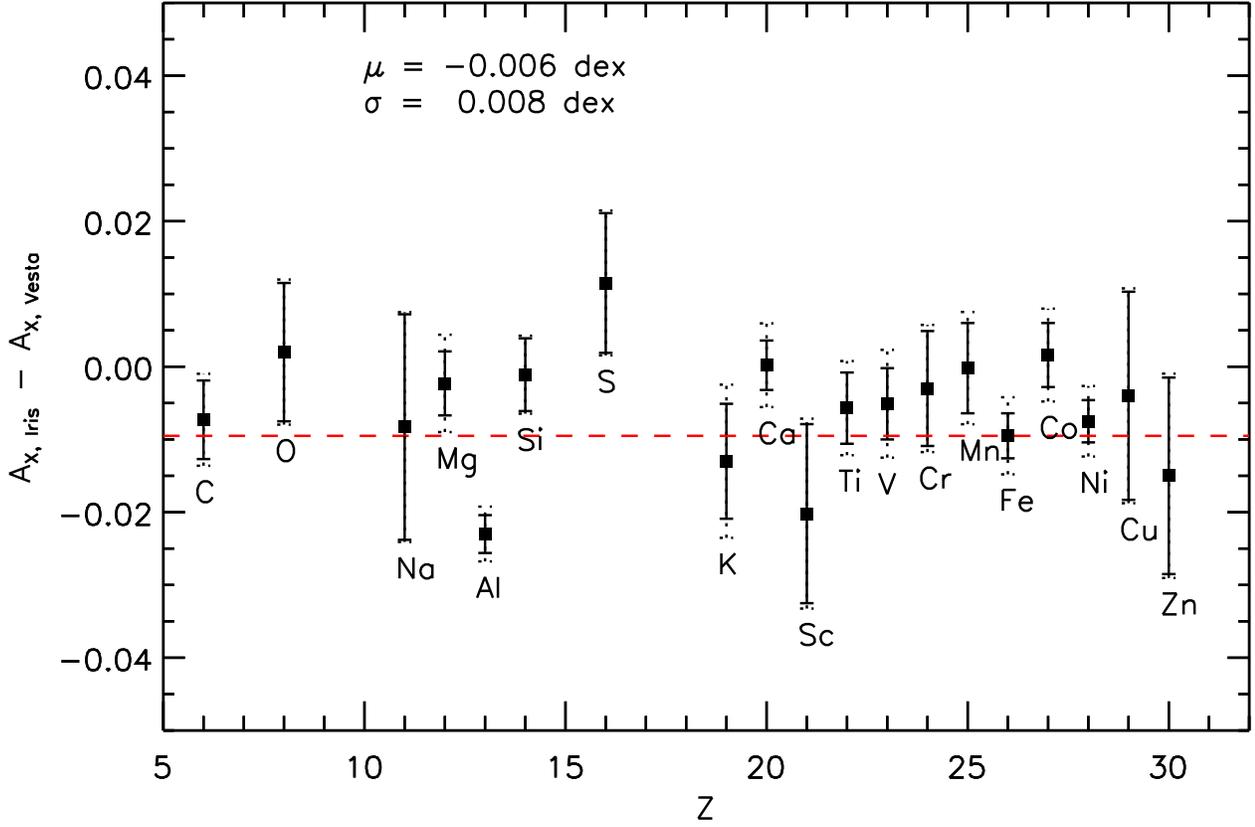}
\caption{Differential abundances for Iris and Vesta data from the MIKE instrument.  Error bars and red line as in Figure \ref{fig:vestamike}.}
\label{fig:irismike}
\end{figure}

An independent test of asteroid-induced errors was performed using the MIKE spectra of Iris and Vesta 1.  These spectra were obtained nearly six months apart in time.  The [X/H] values for Iris with respect to Vesta have a mean and standard deviation consistent with those of MIKE Vesta test on time-dependent systematics (Figure \ref{fig:irismike}, Table \ref{abund_iris}).  Again, the largest outliers from the mean (Al at 4.25$\sigma$ and S at 1.7$\sigma$) are elements for which our line list is somewhat limited: Al has four lines grouped as two doublets, while S has four lines, two of which are a doublet and the other two of which are separated by less than 15 \r{A}.  The strong dependence of our method on a good choice of local pseudo-continuum for the differential line measurements can lead to errors in the abundances from imperfect spectral normalization or unresolved features in the continuum.  These errors will not be fully reflected in the line-to-line scatter if multiple lines come from the same local region and carry the same bias.  This underestimation of the error can be reduced, if not eliminated, by choosing different local pseudo-continua for different lines within the same region; this approach was used for the O triplet and the single K line with good results in these analyses.

Neither of the above tests show any bias on abundances arising from use of different asteroids.  These results are consistent with a past indirect test on the asteroids Ceres and Juno performed by \citet[][Appendix B]{Melendez2012}, which found an element-to-element scatter on abundances of 0.005 dex.

\subsubsection{Choice of Instrument}
\label{mikeesp}

The instrument used for the observation and its line spread function are expected to play a significant role in the spectral analysis when pushing the boundaries of the highest abundance precision.  Although line equivalent widths are in principle independent of resolution, our use of the local pseudo-continuum in EW measurements means that the resolving power applied to the small features surrounding the line in question can make a  difference to the line measurement.  Additionally, fitting each line with a Gaussian will naturally lead to errors on the fit due to the inherent non-Gaussianity of the line and of the spectrograph's instrumental profile.  When measuring lines differentially using the same instrument, the resolution and instrumental profile of each spectrum are roughly the same and these effects cancel out.  If the spectra of the target and reference objects are taken with different instruments, though, the effect of different resolution in the continua could become more important.  Moreover, comparing equivalent widths measured as Gaussians from spectra with substantially different non-Gaussian instrumental profiles could introduce significant errors even for instruments with similar nominal resolutions.

\begin{figure}
\includegraphics[width=\columnwidth]{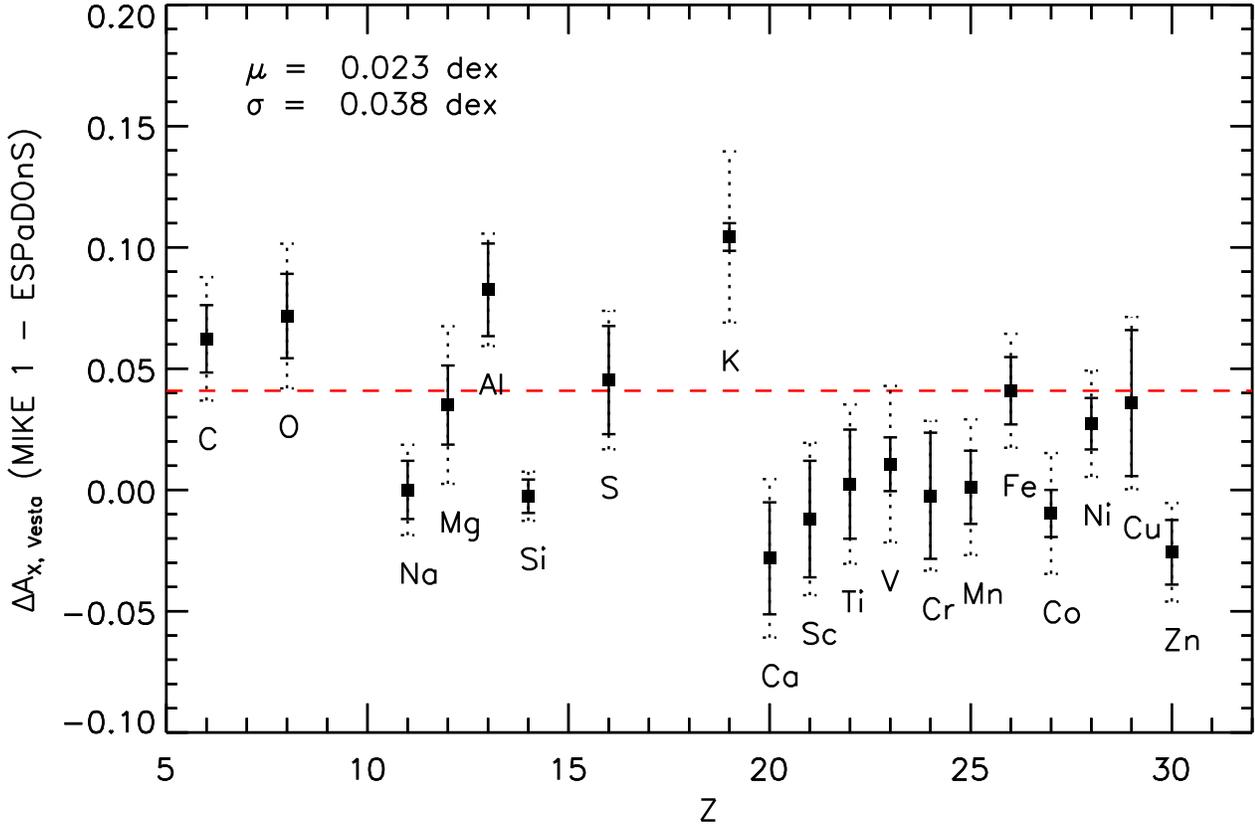}
\caption{Differential abundances for Vesta data from MIKE (taken on 2011 June 24) and ESPaDOnS.  Error bars and red line as in Figure \ref{fig:vestamike}.  Note that the y-axis is rescaled compared to previous figures.}
\label{fig:vestamikemesp}
\end{figure}

To test this effect, we used the ESPaDOnS Vesta spectrum as a reference and the MIKE Vesta 1 spectrum as the target.  We trimmed the line list to exclude any lines not present in both spectra.  Line EW measurements had been performed separately for these two spectra, meaning that the location of the chosen pseudo-continuum likely varied in some cases.  To mitigate this effect, we ran an initial abundance analysis, remeasured the lines which gave the most severe outliers in abundance for each element using a consistent choice of continuum, and re-ran the analysis.  The results have by far the largest errors of any analysis considered in this paper, with a standard deviation of 0.04 dex (Figure \ref{fig:vestamikemesp}, Table \ref{abund_mikemesp}).  Due to the relatively large scatter in Fe line abundances, the stellar parameters are also quite uncertain, with the estimated errors being 5-6 times larger than in previous tests (Table \ref{parameters}).  The relatively poor resulting stellar parameters, in particular the retrieved $log(g)$ of 0.09 $\pm$ 0.06 dex below the nominal solar value and the retrieved metallicity of 0.04 $\pm$ 0.02 dex above the solar value, make model-based systematic errors in the elemental abundances likely.  Abundance precision of 0.01 dex was achievable only in one element, Si.  While a better optimized line list and a differential approach to measuring every line would likely improve the precision to some extent, these results suggest that comparing spectra which were obtained with different instruments is inadvisable for the desired high-precision results.

\subsection{Expected Errors}

In this section, we consider the expected level of error based on formal uncertainties in the data and compare this to the observed error levels.

\subsubsection{Equivalent Width Measurement Uncertainties}

The equivalent widths of the spectral lines are the directly measured quantities in the abundance analysis, so it is key to understand the level of uncertainty in the measured EWs and its effect on the final results.  The expected root-mean-square error in EW based on photon statistics is given in \citet{Cayrel1988} as:

\begin{equation}
\label{eqn:cayrel}
<\sigma_{EW}^2> ^{\frac{1}{2}} \: \simeq \: 1.6 (w \delta x)^{\frac{1}{2}} \epsilon
\end{equation}

\noindent where $w$ is the Gaussian full width at half maximum of the line, $\delta x$ is the pixel size in wavelength units, and $\epsilon$ is the relative photometric accuracy of the continuum, taken here to be the inverse of the local SNR.

For the spectra used in this analysis, typical EW errors from Equation \ref{eqn:cayrel} are on order of 0.1 m\r{A}.  We can estimate the effect of these EW errors on the final abundances by drawing a simulated EW from a Gaussian distribution centered on the measured value and with a $\sigma$ of 0.1 m\r{A} for each line used in the analysis.  Adding random errors in this manner on every line EW for the pair of MIKE Vesta spectra yielded a typical change in each elemental abundance of 0.001 dex.  The statistical error bars on the log abundances increased by only 7 $\pm$ 10 \%.


We conclude from this test that the error on the EW measurements expected from photon statistics generally makes up a small portion of the total statistical errors on the final abundances at high SNR.  Factors such as blending of unresolved lines and continuum deformation from unresolved lines or telluric features are more likely to dominate the errors on measured EWs, causing the level of line-to-line scatter which we observe.  For this reason, high resolution and high SNR are critical factors in obtaining high precision abundances, since they enable the identification of the optimal continuum choice and accurate measurements of line EWs.

\subsubsection{Stellar Parameter Uncertainties}
\label{paramerrors}

\begin{table*}
\caption{Summary of derived sun-as-a-star parameters.}
\label{parameters}
\centering 
\begin{tabular}{l|cccccccc} 
\hline    
\hline 
{Spectrum}& $T_{\textrm{eff}}$ & $\sigma_{T}$ & log $g$ & $\sigma_{logg}$ & $v_t$ & $\sigma_{v_t}$ & [M/H] & $\sigma_{[M/H]}$ \\
{}               & (K)           & (K)                 & (dex)     & (dex)                   & (km s$^{-1}$) & (km s$^{-1}$) & (dex) & (dex)  \\
\hline
Iris (MIKE) \footnotemark[1] & 5769 & 5 & 4.42 & 0.01 & 0.86 &  0.01 & -0.01 & 0.01 \\
Vesta (MIKE, 2) \footnotemark[1] & 5779 & 5 & 4.44 & 0.01 & 0.86 & 0.01 & -0.01 & 0.01 \\
Ceres (ESPaDOnS) \footnotemark[2] & 5778 & 8 & 4.44 & 0.02 & 0.85 & 0.01 & 0.00 & 0.01 \\
Vesta (MIKE, 1) \footnotemark[2] & 5780 & 29 & 4.35 & 0.06 & 0.87 & 0.05 & 0.04 & 0.02 \\
\hline       
\end{tabular}
\footnotetext[1]{Measured differentially with respect to standard MIKE Vesta 1, with assumed solar parameters ($T_{\textrm{eff}}$ = 5777 K, log $g$ = 4.44, $v_t$ = 0.85 km s$^{-1}$, [M/H] = 0.00).}
\footnotetext[2]{Measured differentially with respect to standard ESPaDOnS Vesta, with assumed solar parameters as above.}

\end{table*}

The uncertainty in physical parameters of the target star can lead to use of a sub-optimal model atmosphere in the abundance analysis, creating additional errors.  Uncertainties on each parameter were propagated to individual abundances by running the MOOG \textit{abfind} analysis with model atmospheres which varied each parameter by its one-sigma error bar, and the resulting abundance changes due to each parameter were added in quadrature to yield a net ``parameter-based uncertainty" on every abundance.  

In general, the parameter-based uncertainties are at or below the level of statistical uncertainty inferred from the line-to-line scatter (Tables \ref{abund_vestamike} - \ref{abund_iris}).  This implies that with high spectral quality and a sufficiently long and balanced Fe line list, which enable stellar parameter precisions on the level of those achieved in our tests, the derived stellar parameters are not the limiting factor on abundance precision.  A notable exception is the case of Vesta MIKE - ESPaDOnS (Table \ref{abund_mikemesp}), where the much larger uncertainties on the parameters cause the parameter-based uncertainties to dominate the errors on abundances.  As discussed in section \ref{mikeesp}, the additional errors introduced by the use of different instruments led to a larger scatter in the derived Fe abundances, making it difficult to achieve parameters as precise as those found in the other tests.


\section{Implications for Planet Signatures}

\begin{figure}
\includegraphics[width=\columnwidth]{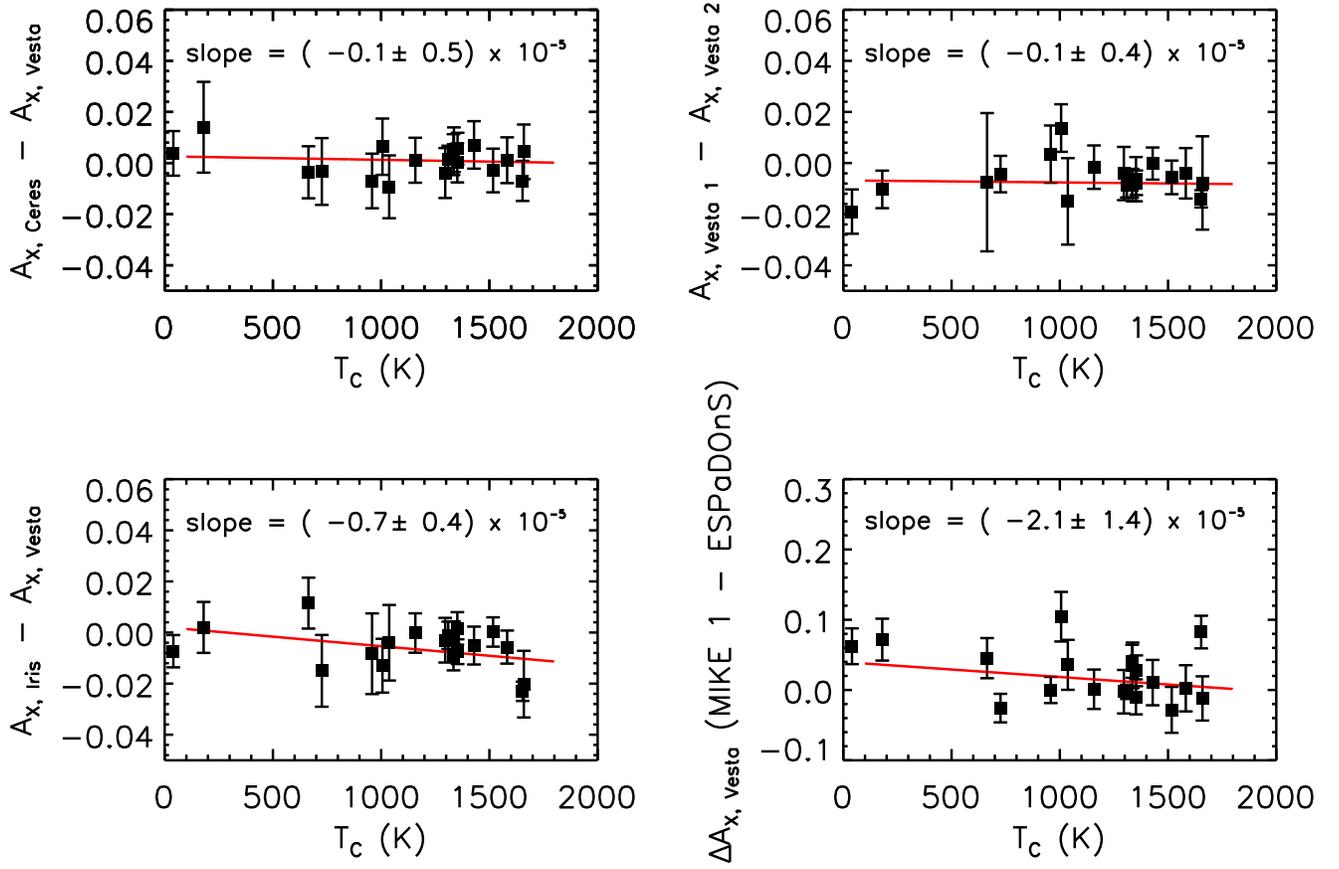}
\caption{Differential abundances plotted against the 50 \% condensation temperature from \citet{Lodders2003}.  Linear fits to the data were performed using the total error bars as weights and are shown in red.}
\label{fig:tcond}
\end{figure}

One past result that critically depends on high-precision abundances is the Sun's evident chemical depletion trend with condensation temperature relative to the average solar twin star \citep{Melendez2009}, often interpreted as a potential sign of past terrestrial planet formation.  We searched for trends in elemental abundance with the 50\% condensation temperature from \citet{Lodders2003} in each target spectrum considered above.  All solar spectra were found to have a slope in abundance vs. condensation temperature consistent with zero within 2$\sigma$ (Figure \ref{fig:tcond}).  It is important to note, however, that characterizing the stellar parameter uncertainties correctly is a key part of evaluating the significance of a potential slope with condensation temperature.  The elements at low condensation temperature, especially C and O, are derived from high excitation potential lines.  An incorrect stellar effective temperature can change the abundances of C and O relative to the refractory elements and induce a false slope.  One way to mitigate this problem is to derive the carbon abundance separately from CH molecular lines.  This method was not used in this paper due to the CH lines under consideration falling outside of the MIKE spectral range.

We vary the model parameters from each best fit solution in order to estimate the error in temperature needed to reproduce an abundance slope of the size typically investigated as potential planet formation signatures, as in e.g. \citet{Schuler2011} or \citet{Ramirez2013}.  To produce a spurious slope of $5 \times 10^{-5}$ dex K$^{-1}$, the effective temperature chosen would need to be around 60-70 K below the true value.  For this slope to be a statistically significant result, of course, the formal uncertainties on the stellar parameters would need to be far below the level of the true error.  For this reason, it is critical to evaluate the stellar effective temperature and its error completely, preferably using multiple methods of temperature determination.  In most of the analyses carried out in this paper, the solar temperature is retrieved for the target with an error well below the estimated 1$\sigma$ error bar.


\section{Conclusion}

Based on the tests conducted on solar spectra taken with different asteroids as reflectors, different instruments, and different epochs of observation, we conclude that precision below the level of 0.01 dex is achievable given high-quality target and reference spectra obtained with the same instrument.  Time-dependent effects on the scale of several months appear relatively unimportant, as does the choice of asteroid used for a reflected solar spectrum.  Due to the need for a constant pseudo-continuum level for measuring differential equivalent widths, the line spread function of the target spectrum relative to the reference spectrum is a critical factor in measurement precision.  This effect means that using target and reference spectra taken with different instruments is inadvisable.

The greatest contributor to the achievable precision of an individual element is the line list employed in the analysis.  For the most part this effect can be quantified by using the standard error on the mean abundance as a statistical error bar, but we urge caution when applying this method of error estimation to elements which have multiple lines within the same small wavelength region.  As seen by the examples of K and Al in these analyses, the statistical error can underestimate potential effects from local continuum normalization or unresolved blends on the consistency of the pseudo-continuum across spectra.

Past work using this technique of high-precision differential abundance analysis has demonstrated that the results are free of potential systematic biases at or above the level of 0.01 dex originating from the model atmosphere used \citep{Ramirez2011,Melendez2012} or from non-LTE effects \citep{Melendez2012}.  We can now conclude that time-dependent instrumental effects and the choice of asteroid for the solar standard are also free of such errors.  Given a thorough understanding of the limitations of one's line list and use of the same instrument for the target and reference spectra, we find no reason to doubt the reality of sub-0.01 dex precisions on differential abundances for spectrally similar stars.

\acknowledgements{M.B. is supported by a National Science Foundation Graduate Research Fellowship under Grant No. DGE-1144082.  J.M. thanks FAPESP (2012/24392-2).  J.L.B. acknowledges support for this work from the NSF (grant number AST-1313119) and the Alfred P. Sloan Foundation.}

{\it Facilities:} \facility{Magellan:Clay (MIKE)}, \facility{CFHT (ESPaDOnS)}.

\pagebreak
\bibliographystyle{apj}
\bibliography{asteroids_ms}

\pagebreak

\pagebreak

\begin{table}
\caption{Stellar abundances [X/H] for MIKE Vesta 2 - Vesta 1.}
\label{abund_vestamike}
\centering 
\begin{tabular}{lccccccccc} 
\hline    
\hline 
{Element}& [X/H]$_{Vesta}$   & $\Delta T_{\textrm{eff}}$ & $\Delta$log $g$ & $\Delta v_t$ & $\Delta$[M/H] & param\footnotemark[1] & obs\footnotemark[2]  & total \\
{}       &       & +5K           &  +0.01 dex      & +0.01 km s$^{-1}$  & +0.01 dex   &  &  &  \\
{}       & (dex) & (dex)         & (dex)           & (dex)       & (dex)        & (dex) & (dex) & (dex) \\
\hline
C &  -0.019 &  -0.003 &   0.002 &   0.000 &  -0.001 &   0.003 &   0.008 &   0.009 \\
O &  -0.010 &  -0.003 &   0.000 &   0.000 &   0.000 &   0.003 &   0.007 &   0.007 \\
Na &   0.004 &   0.002 &   0.000 &  -0.000 &   0.001 &   0.002 &   0.011 &   0.011 \\
Mg &  -0.009 &   0.004 &  -0.001 &  -0.002 &   0.001 &   0.005 &   0.004 &   0.006 \\
Al &  -0.014 &   0.003 &  -0.001 &   0.000 &   0.001 &   0.003 &   0.002 &   0.003 \\
Si &  -0.009 &   0.001 &   0.001 &  -0.000 &   0.001 &   0.002 &   0.004 &   0.005 \\
S &  -0.007 &  -0.002 &   0.002 &   0.000 &   0.000 &   0.003 &   0.027 &   0.027 \\
K &   0.014 &   0.005 &  -0.003 &  -0.001 &   0.003 &   0.007 &   0.006 &   0.009 \\
Ca &  -0.006 &   0.003 &  -0.002 &  -0.002 &   0.002 &   0.004 &   0.005 &   0.007 \\
Sc\footnotemark[3] &  -0.008 &   0.002 &   0.002 &  -0.001 &   0.002 &   0.004 &   0.018 &   0.018 \\
Ti\footnotemark[3] &  -0.004 &   0.004 &   0.001 &  -0.001 &   0.002 &   0.005 &   0.009 &   0.010 \\
V &  -0.000 &   0.005 &   0.001 &  -0.000 &   0.001 &   0.005 &   0.003 &   0.006 \\
Cr\footnotemark[3] &  -0.004 &   0.003 &   0.000 &  -0.002 &   0.002 &   0.004 &   0.010 &   0.010 \\
Mn &  -0.002 &   0.004 &  -0.001 &  -0.002 &   0.001 &   0.005 &   0.007 &   0.009 \\
Co &  -0.006 &   0.004 &   0.001 &  -0.000 &   0.002 &   0.004 &   0.007 &   0.009 \\
Ni &  -0.008 &   0.003 &   0.000 &  -0.001 &   0.002 &   0.004 &   0.003 &   0.005 \\
Cu &  -0.015 &   0.003 &   0.001 &  -0.001 &   0.002 &   0.005 &   0.016 &   0.017 \\
Zn &  -0.004 &   0.001 &   0.001 &  -0.002 &   0.003 &   0.004 &   0.006 &   0.007 \\
Fe\footnotemark[3] &  -0.008 &   0.003 &  -0.000 &  -0.002 &   0.002 &   0.004 &   0.004 &   0.006 \\

\hline       
\end{tabular}
\footnotetext[1]{Error due to propagating formal uncertainties on the stellar parameters.}
\footnotetext[2]{Statistical error reflected in the line-to-line scatter of derived abundances.}
\footnotetext[3]{Abundances and errors quoted are the weighted mean of the results for two ionization states.}
\end{table}

\begin{table}
\caption{Stellar abundances [X/H] for ESPaDOnS Ceres - Vesta.}
\label{abund_esp}
\centering 
\begin{tabular}{lccccccccc} 
\hline    
\hline 
{Element}& [X/H]$_{Ceres}$ & $\Delta T_{\textrm{eff}}$ & $\Delta$log $g$ & $\Delta v_t$ & $\Delta$[M/H] & param\footnotemark[1] & obs\footnotemark[2] & total \\
{}       &       &  +8K           &  +0.02 dex      & +0.01 km s$^{-1}$  & +0.01 dex   &  &  &  \\
{}       & (dex) & (dex)         & (dex)           & (dex)       & (dex)        & (dex) & (dex) & (dex) \\
\hline
C &   0.004 &  -0.004 &   0.004 &   0.000 &   0.001 &   0.006 &   0.007 &   0.009 \\
O &   0.014 &  -0.006 &   0.000 &  -0.001 &   0.002 &   0.006 &   0.017 &   0.018 \\
Na &  -0.007 &   0.004 &   0.000 &  -0.000 &  -0.000 &   0.004 &   0.010 &   0.011 \\
Mg &   0.005 &   0.005 &  -0.001 &  -0.001 &  -0.001 &   0.006 &   0.007 &   0.009 \\
Al &  -0.007 &   0.003 &  -0.002 &  -0.000 &  -0.000 &   0.004 &   0.007 &   0.008 \\
Si &   0.001 &   0.002 &   0.001 &  -0.000 &   0.001 &   0.002 &   0.005 &   0.005 \\
S &  -0.004 &  -0.004 &   0.003 &  -0.000 &   0.001 &   0.005 &   0.009 &   0.010 \\
K &   0.006 &   0.006 &  -0.006 &  -0.002 &   0.002 &   0.009 &   0.006 &   0.011 \\
Ca &  -0.003 &   0.006 &  -0.003 &  -0.002 &   0.000 &   0.006 &   0.006 &   0.009 \\
Sc\footnotemark[3] &   0.004 &   0.002 &   0.006 &  -0.002 &   0.002 &   0.006 &   0.009 &   0.011 \\
Ti\footnotemark[3] &   0.001 &   0.004 &   0.004 &  -0.002 &   0.001 &   0.006 &   0.006 &   0.009 \\
V &   0.007 &   0.008 &   0.002 &  -0.000 &  -0.001 &   0.008 &   0.004 &   0.009 \\
Cr\footnotemark[3] &  -0.004 &   0.003 &   0.003 &  -0.002 &   0.001 &   0.005 &   0.009 &   0.010 \\
Mn &   0.001 &   0.006 &  -0.000 &  -0.001 &  -0.000 &   0.006 &   0.006 &   0.009 \\
Co &   0.000 &   0.006 &   0.003 &  -0.000 &   0.000 &   0.007 &   0.003 &   0.008 \\
Ni &   0.006 &   0.005 &   0.000 &  -0.002 &   0.001 &   0.005 &   0.004 &   0.006 \\
Cu &  -0.009 &   0.003 &   0.001 &  -0.001 &   0.002 &   0.004 &   0.011 &   0.012 \\
Zn &  -0.003 &   0.002 &   0.001 &  -0.002 &   0.002 &   0.004 &   0.013 &   0.013 \\
Fe\footnotemark[3] &   0.003 &   0.006 &   0.000 &  -0.002 &   0.001 &   0.006 &   0.005 &   0.008 \\
\hline       
\end{tabular}
\footnotetext[1]{Error due to propagating formal uncertainties on the stellar parameters.}
\footnotetext[2]{Statistical error reflected in the line-to-line scatter of derived abundances.}
\footnotetext[3]{Abundances and errors quoted are the weighted mean of the results for two ionization states.}
\end{table}

\begin{table}
\caption{Stellar abundances [X/H] for MIKE Iris - Vesta.}
\label{abund_iris}
\centering 
\begin{tabular}{lccccccccc} 
\hline    
\hline 
{Element}& [X/H]$_{Iris}$   & $\Delta T_{\textrm{eff}}$ & $\Delta$log $g$ & $\Delta v_t$ & $\Delta$[M/H] & param\footnotemark[1] & obs\footnotemark[2]  & total \\
{}       &       & +5K           &  +0.01 dex      & +0.01 km s$^{-1}$  & +0.01 dex   &  &  &  \\
{}       & (dex) & (dex)         & (dex)           & (dex)       & (dex)        & (dex) & (dex) & (dex) \\
\hline
C &  -0.008 &  -0.002 &   0.002 &   0.000 &  -0.001 &   0.003 &   0.007 &   0.008 \\
O &   0.002 &  -0.003 &   0.000 &   0.000 &   0.000 &   0.003 &   0.010 &   0.011 \\
Na &  -0.008 &   0.003 &   0.000 &  -0.000 &   0.001 &   0.003 &   0.015 &   0.016 \\
Mg &  -0.002 &   0.004 &  -0.001 &  -0.002 &   0.001 &   0.005 &   0.004 &   0.007 \\
Al &  -0.023 &   0.002 &  -0.001 &  -0.001 &   0.001 &   0.003 &   0.003 &   0.004 \\
Si &  -0.001 &   0.001 &   0.001 &  -0.000 &   0.001 &   0.002 &   0.005 &   0.005 \\
S &   0.011 &  -0.002 &   0.002 &   0.000 &  -0.000 &   0.003 &   0.010 &   0.010 \\
K &  -0.013 &   0.005 &  -0.003 &  -0.001 &   0.004 &   0.007 &   0.008 &   0.011 \\
Ca &   0.000 &   0.004 &  -0.002 &  -0.002 &   0.001 &   0.005 &   0.003 &   0.006 \\
Sc\footnotemark[3] &  -0.020 &   0.002 &   0.002 &  -0.001 &   0.002 &   0.004 &   0.012 &   0.013 \\
Ti\footnotemark[3] &  -0.006 &   0.002 &   0.002 &  -0.002 &   0.002 &   0.004 &   0.005 &   0.006 \\
V &  -0.005 &   0.005 &   0.001 &  -0.000 &   0.001 &   0.006 &   0.005 &   0.008 \\
Cr\footnotemark[3] &  -0.003 &   0.003 &   0.000 &  -0.002 &   0.002 &   0.004 &   0.008 &   0.009 \\
Mn &  -0.000 &   0.004 &  -0.002 &  -0.003 &   0.002 &   0.005 &   0.006 &   0.008 \\
Co &   0.002 &   0.004 &   0.001 &  -0.001 &   0.002 &   0.004 &   0.005 &   0.007 \\
Ni &  -0.007 &   0.003 &  -0.000 &  -0.002 &   0.002 &   0.004 &   0.003 &   0.005 \\
Cu &  -0.003 &   0.003 &  -0.001 &  -0.002 &   0.002 &   0.004 &   0.015 &   0.015 \\
Zn &  -0.015 &   0.001 &   0.000 &  -0.003 &   0.002 &   0.004 &   0.014 &   0.014 \\
Fe\footnotemark[3] &  -0.009 &   0.003 &   0.000 &  -0.002 &   0.002 &   0.004 &   0.003 &   0.005 \\
\hline       
\end{tabular}
\footnotetext[1]{Error due to propagating formal uncertainties on the stellar parameters.}
\footnotetext[2]{Statistical error reflected in the line-to-line scatter of derived abundances.}
\footnotetext[3]{Abundances and errors quoted are the weighted mean of the results for two ionization states.}
\end{table}

\begin{table}
\caption{Stellar abundances [X/H] for Vesta MIKE 1 - ESPaDoNS.}
\label{abund_mikemesp}
\centering 
\begin{tabular}{lccccccccc} 
\hline    
\hline 
{Element}& [X/H]$_{Vesta}$   & $\Delta T_{\textrm{eff}}$ & $\Delta$log $g$ & $\Delta v_t$ & $\Delta$[M/H] & param\footnotemark[1] & obs\footnotemark[2]  & total \\
{}       &       & +29K           &  +0.06 dex      & +0.05 km s$^{-1}$  & +0.02 dex   &  &  &  \\
{}       & (dex) & (dex)         & (dex)           & (dex)       & (dex)        & (dex) & (dex) & (dex) \\
\hline
C &   0.062 &  -0.016 &   0.014 &  -0.001 &   0.001 &   0.021 &   0.014 &   0.025 \\
O &   0.072 &  -0.023 &   0.006 &  -0.003 &   0.004 &   0.024 &   0.017 &   0.030 \\
Na &   0.000 &   0.014 &  -0.001 &  -0.001 &  -0.000 &   0.014 &   0.012 &   0.019 \\
Mg &   0.035 &   0.025 &  -0.007 &  -0.010 &  -0.001 &   0.028 &   0.016 &   0.033 \\
Al &   0.083 &   0.012 &  -0.004 &  -0.002 &  -0.000 &   0.013 &   0.019 &   0.023 \\
Si &  -0.003 &   0.006 &   0.003 &  -0.002 &   0.002 &   0.007 &   0.007 &   0.010 \\
S &   0.045 &  -0.013 &   0.012 &  -0.001 &   0.002 &   0.018 &   0.022 &   0.029 \\
K &   0.104 &   0.025 &  -0.022 &  -0.009 &   0.003 &   0.035 &   0.006 &   0.035 \\
Ca &  -0.028 &   0.019 &  -0.010 &  -0.008 &   0.001 &   0.023 &   0.023 &   0.033 \\
Sc\footnotemark[3] &  -0.012 &   0.006 &   0.017 &  -0.007 &   0.005 &   0.020 &   0.024 &   0.031 \\
Ti\footnotemark[3] &   0.002 &   0.022 &   0.006 &  -0.006 &   0.000 &   0.024 &   0.022 &   0.033 \\
V &   0.011 &   0.030 &   0.003 &  -0.002 &  -0.001 &   0.030 &   0.011 &   0.032 \\
Cr\footnotemark[3] &  -0.002 &   0.015 &   0.003 &  -0.007 &   0.001 &   0.017 &   0.026 &   0.031 \\
Mn &   0.001 &   0.023 &  -0.002 &  -0.006 &  -0.000 &   0.024 &   0.015 &   0.028 \\
Co &  -0.010 &   0.022 &   0.006 &  -0.001 &   0.000 &   0.023 &   0.010 &   0.025 \\
Ni &   0.027 &   0.017 &  -0.001 &  -0.009 &   0.002 &   0.019 &   0.011 &   0.022 \\
Cu &   0.036 &   0.017 &   0.002 &  -0.009 &   0.001 &   0.019 &   0.030 &   0.036 \\
Zn &  -0.026 &   0.004 &   0.004 &  -0.013 &   0.005 &   0.015 &   0.013 &   0.020 \\
Fe\footnotemark[3] &   0.041 &   0.015 &   0.002 &  -0.011 &   0.002 &   0.019 &   0.014 &   0.023 \\

\hline       
\end{tabular}
\footnotetext[1]{Error due to propagating formal uncertainties on the stellar parameters.}
\footnotetext[2]{Statistical error reflected in the line-to-line scatter of derived abundances.}
\footnotetext[3]{Abundances and errors quoted are the weighted mean of the results for two ionization states.}
\end{table}

\end{document}